\documentclass[11pt,superscriptaddress,aps,prd,preprint,showpacs]{revtex4}
\usepackage[dvips]{graphicx}
\usepackage{amsfonts}
\usepackage{slashed}
\usepackage{amsmath}
\usepackage{slashed}
\newcommand{\be}{\begin{equation}}
\newcommand{\en}{\end{equation}}
\newcommand{\ee}{\end{equation}}
\newcommand{\ben}{\begin{eqnarray}}
\newcommand{\een}{\end{eqnarray}}
\newcommand{\RM}[1]{\mathrm{#1}}

\newcommand{\remark}[1]{}

\begin{document}

\title{Induced Chern-Simons modified gravity at finite temperature}

\author{J. F. Assun\c c\~ao}
\author{T. Mariz}
\affiliation{Instituto de F\'\i sica, Universidade Federal de Alagoas,\\ 
57072-900, Macei\'o, Alagoas, Brazil}
\email{jfassuncao, tmariz@fis.ufal.br}

\author{J. R. Nascimento}
\author{A. Yu. Petrov}
\affiliation{Departamento de F\'{\i}sica, Universidade Federal da Para\'\i ba,\\
Caixa Postal 5008, 58051-970, Jo\~ao Pessoa, Para\'\i ba, Brazil}
\email{jroberto, petrov@fisica.ufpb.br}

%\date{\today}

\begin{abstract}
We calculate the linearized four-dimensional gravitational Chern-Simons term at the finite temperature, show its finiteness and explicitly demonstrate that its transversal part matches the known result for the chiral vortical conductivity.
\end{abstract}


\maketitle

As it is known, the Lorentz and CPT symmetries are violated \cite{Colladay:1996iz,Colladay:1998fq,Kostelecky:2003fs} when we add the four-dimensional gravitational Chern-Simons term, in the weak (linearized) gravity case, looking like ${\cal L}_b=\frac14b^\lambda h^{\mu\nu}\epsilon_{\alpha\mu\lambda\rho}\partial^\rho(\partial_\gamma\partial^\gamma h^\alpha_\nu-\partial_\nu\partial_\gamma h^{\gamma\alpha})$, to the Einstein-Hilbert Lagrangian \cite{Jackiw:2003pm,Mariz:2004cv,Mariz:2007gf,Gomes:2008an}. This term has been shown to arise as a quantum correction in a theory describing coupling of gravity to fermions \cite{Mariz:2004cv,Mariz:2007gf,Gomes:2008an}, and recently, it has been shown \cite{Gomes:2008an} that this term displays the ambiguity, similar to that one of the Carroll-Field-Jackiw term ${\cal L}_{CFJ}=\frac12b_\mu\epsilon^{\mu\nu\lambda\rho}F_{\nu\lambda}A_\rho$ \cite{Jackiw:1999yp,Chung:1999gg,Jackiw:1999qq}.

In this work we have interested in extending the generation of the gravitational Chern-Simons term to the finite temperature case. The main motivation for this study is the interest to the anomalies in the curved spacetime, initially inspired by \cite{Son:2009tf}, where the triangle anomaly of the gauge field has been studied in the curved spacetime at the finite temperature. In fact, here we meet the appearance of new non-dissipative energy transport phenomena, observed in relativistic hydrodynamics \cite{Landsteiner:2012kd}, given by $\vec{J}_\epsilon = \sigma \vec{\omega}$, where $\vec{J}_\epsilon$ is the energy current, $\sigma$ is the transport coefficient (conductivity), and $\vec{\omega}$ is a vector or pseudo-vector inducing the transport. In this scenario, an important study has been carried out in \cite{Chernodub:2013kya}, where the emergence of the energy current $\vec{J}_\epsilon$ as a consequence of mixed gauge and gravitational fields, in a Weyl semimetal system, was claimed as a manifestation of the axial magnetic effect (AME), with $\omega_{i} = \frac12 \epsilon_{ijk}\partial_{j}A_{k}$ being the axial magnetic field and $\sigma=\sigma_{AME}$ the temperature-dependent conductivity. Thus, the aim we pursue in this paper is the calculation of another temperature-dependent contribution for the same energy current, due only to the gravitational fields, through the chiral vortical effect (CVE) \cite{Erdmenger:2008rm,Banerjee:2008th}, where now $\omega_{i} = \frac12 \epsilon_{ijk}\partial_{j}h_{0k}$ is generated by the metric fluctuation $h_{\mu\nu}$.

Our starting point is the fermionic action (see f.e. \cite{Mariz:2004cv}) given by
\be
S_\psi = \int d^4x\,e\,\left(\frac i2{e^\mu}_a\bar\psi\gamma^a {\stackrel{\leftrightarrow}{D_\mu}} \psi - \,m\bar\psi\psi-b_\mu{e^\mu}_a\bar\psi\gamma^a\gamma_5\psi\right),
\ee
where $e^{\mu}_{\phantom{a}a}$ is the tetrad, $e\equiv{\rm det}\,e^{\mu}_{\phantom{a}a}$, and $D_{\mu}\psi=\partial_{\mu}\psi-i\omega_{\mu}\psi$ ($D_{\mu}\bar\psi=\partial_{\mu}\bar\psi+i\omega_{\mu}\bar\psi$), with $\omega_\mu=\frac{1}{4}\omega_{\mu bc}\sigma^{bc}$ being the spin connection, and $\sigma^{bc}=\frac{i}{2}[\gamma^b,\gamma^c]$. Note that, in the term $b_\mu{e^\mu}_a\bar\psi\gamma^a\gamma_5\psi$, we have the Lorentz-violating coefficient $b_\mu$ and the CPT-violating operator ${e^\mu}_a\bar\psi\gamma^a\gamma_5\psi$.

In order to obtain the effective action, we must consider the fermionic generating functional
\begin{eqnarray}
Z_\psi &=& \int D\psi D\bar\psi e^{iS_\psi}=e^{iS_\RM{eff}},
\end{eqnarray}
so that after we perform the fermionic integration, we get
\be
S_\RM{eff} = -i\RM{Tr}\ln\left(\frac i2e\,{e^\mu}_a\gamma^a {\stackrel{\leftrightarrow}{\partial_\mu}} - e\,m - e\,b_\mu {e^\mu}_a\gamma^a\gamma_5 + e\,{e^\mu}_a\gamma^a\omega_\mu \right),
\en
where $\RM{Tr}$ stands for the trace over Dirac matrices as well as for the functional trace corresponding to the integration in momentum and coordinate spaces.

Throughout this paper, we use the weak field approximation, in which the tetrad and the
connection are expressed in terms of the metric fluctuation $h_{\mu\nu}$ as $e_{\mu a}=\eta_{\mu a}+\frac{1}{2}h_{\mu a}$ and 
$\omega_{\mu ab}=\frac{1}{2}(\partial_b h_{\mu a}-\partial_a h_{\mu b})$ \cite{Choi:1994ax}. Thus, we have
\be\label{Seffh}
S_\RM{eff} = -i\RM{Tr}\ln\left[i\slashed{\partial} - m - \slashed{b}\gamma_5 - \frac i4 h_{\mu\nu}\gamma^\mu {\stackrel{\leftrightarrow}{\partial^\nu}} + \frac i{32}(h_{\mu\nu} {\stackrel{\leftrightarrow}{\partial_\lambda}} {h_\rho}^\nu)\gamma^{\lambda}\gamma^{\rho}\gamma^{\mu}\right].
\en

Then, in order to single out the quadratic terms in $h_{\mu\nu}$ within the effective action, we rewrite the expression (\ref{Seffh}) as $S_\RM{eff}=S_\RM{eff}^{(0)}+\sum\limits_{n=1}^\infty S_\RM{eff}^{(n)}$, where the contribution of an arbitrary $n$-th order in the metric fluctuation yields
\be\label{Seffh2}
S_\RM{eff}^{(n)} = i\RM{Tr}\frac 1n \left\{\frac 1{i\slashed{\partial} - m - \slashed{b}\gamma_5} \left[\frac i4 h_{\mu\nu}\gamma^\mu {\stackrel{\leftrightarrow}{\partial^\nu}} - \frac i{32}(h_{\mu\nu} {\stackrel{\leftrightarrow}{\partial_\lambda}} {h_\rho}^\nu)\gamma^{\lambda}\gamma^{\rho}\gamma^{\mu}\right]\right\}^n.
\en

As our goal is the generation of the gravitational Chern-Simons action, we will single out above terms of the second order in $h_{\mu\nu}$ and first order in $b_\mu$. Firstly, let us analyze the term coming from $n=1$, given by
\be
S_\RM{CS}^{(1)} = -i\RM{Tr}\frac 1{i\slashed{\partial} - m}\slashed{b}\gamma_5\frac 1{i\slashed{\partial} - m} \,\frac i{32}(h_{\mu\nu} {\stackrel{\leftrightarrow}{\partial_\lambda}} {h_\rho}^\nu)\gamma^{\lambda}\gamma^{\rho}\gamma^{\mu}.
\en
Then, after we carry out the traces over the integration in spaces, we obtain
\be\label{CS1}
S_\RM{CS}^{(1)}[h] = i\int d^4x\,h_{\mu\nu}\,\Pi_a^{\mu\nu\rho\sigma}\,h_{\rho\sigma},
\en
with
\be\label{inta}
\Pi_a^{\mu\nu\rho\sigma} = -\frac{1}{16} \RM{tr} \int\frac{d^4p}{(2\pi)^4} S(p)\slashed{b}\gamma_5S(p)\slashed{\partial}\gamma^{\rho}\gamma^{\mu}\eta^{\nu\sigma},
\en
where the symbol $\RM{tr}$ means that the trace is only over Dirac matrices and $S(p)=(\slashed{p}-m)^{-1}$. 

Now, let us consider the terms coming from $n=2$, given by
\ben
S_\RM{CS}^{(2)} &=& \frac i2 \RM{Tr}\frac 1{i\slashed{\partial} - m}\slashed{b}\gamma_5\frac 1{i\slashed{\partial} - m} \frac i4 h_{\mu\nu}\gamma^\mu\!\!\!\!\stackrel{\leftrightarrow}{\,\,\partial\,^\nu}\!\! \frac 1{i\slashed{\partial} - m}\frac i4 h_{\alpha\beta}\gamma^\alpha\!\!\!\!\stackrel{\leftrightarrow}{\,\,\partial\,^\beta}\!\! \nonumber\\
&& + \frac i2 \RM{Tr}\frac 1{i\slashed{\partial} - m} \frac i4 h_{\mu\nu}\gamma^\mu\!\!\!\!\stackrel{\leftrightarrow}{\,\,\partial\,^\nu}\!\! \frac 1{i\slashed{\partial} - m}\slashed{b}\gamma_5\frac 1{i\slashed{\partial} - m} \frac i4 h_{\alpha\beta}\gamma^\alpha\!\!\!\!\stackrel{\leftrightarrow}{\,\,\partial\,^\beta}\!\!. 
\een
By using the key identity of the derivative expansion approach $h_{\mu\nu}(x)S(p)=S(p-i\partial)h_{\mu\nu}(x)$, in order to disentangle the traces over $x_\mu$ and $p_\mu$, we arrive at
\be\label{CS2}
S_\RM{CS}^{(2)}[h] = \frac i2\int d^4x \,h_{\mu\nu}\left(\Pi_b^{\mu\nu\rho\sigma}+\Pi_c^{\mu\nu\rho\sigma}\right)h_{\rho\sigma},
\en
where
\be\label{intb}
\Pi_b^{\mu\nu\rho\sigma} = \frac{1}{16}\RM{tr} \int\frac{d^4p}{(2\pi)^4} S(p) \slashed{b} \gamma_5 S(p) \gamma^\mu (2p^\nu-i\partial^\nu) S(p-i\partial) \gamma^\rho (2p^\sigma-i\partial^\sigma)
\en
and
\be\label{intc}
\Pi_c^{\mu\nu\rho\sigma} = \frac{1}{16}\RM{tr} \int\frac{d^4p}{(2\pi)^4} S(p) \gamma^\mu (2p^\nu-i\partial^\nu) S(p-i\partial)\slashed{b} \gamma_5 S(p-i\partial) \gamma^\rho (2p^\sigma-i\partial^\sigma).
\en

By considering the whole expression for the vacuum polarization tensor, given by $\Pi^{\mu\nu\rho\sigma}=\Pi_a^{\mu\nu\rho\sigma}+\frac12\Pi_b^{\mu\nu\rho\sigma}+\frac12\Pi_c^{\mu\nu\rho\sigma}$,  with $i\partial_\mu \to k_\mu$, we obtain
\begin{eqnarray}\label{PG}
\Pi^{\mu\nu\rho\sigma} &=& \frac{1}{192 \pi ^2}\left[1+\frac{12 m^2}{k^2} \sqrt{\frac{4 m^2}{k^2}-1}\, \csc ^{-1}\left(\frac{2 m}{\sqrt{k^2}}\right)-\frac{12 m^2}{k^2} \right] \left(k^2 \eta^{\nu \sigma}-k^\nu k^\sigma \right)\epsilon ^{\mu \rho \kappa \lambda} b_\kappa k_\lambda,\,\,\,
\end{eqnarray}
where we have first calculated the trace over the Dirac matrices and afterwards, integrated over the momentum $p_\mu$ and parameter $x$ of the Feynman parametrization. For more details, see \cite{Ferrari:2006gs} for the tensor $\Pi_a^{\mu\nu\rho\sigma}$ and \cite{Mariz:2004cv} for the tensors $\Pi_b^{\mu\nu\rho\sigma}$ and $\Pi_c^{\mu\nu\rho\sigma}$. In these works \cite{Mariz:2004cv,Ferrari:2006gs} it was argued that the divergent term disappears when the zero mass limit is taken. However, now, we are observing that the divergent contributions cancel each other, so that, finally, we have the general gauge-invariant expression (\ref{PG}). 

By looking again at Eq. (\ref{PG}), we can easily analyze the limits $k^2\ll m^2$ ($m\neq0$) and $k^2\gg m^2$ ($m=0$), so that we get 
\begin{eqnarray}
\Pi^{\mu\nu\rho\sigma} &=& {\cal O}\left(\frac{k^2}{m^2}\right)
\end{eqnarray}
and
\begin{eqnarray}\label{Pim0}
\Pi^{\mu\nu\rho\sigma} &=& \frac{1}{192 \pi ^2} \left(k^2 \eta^{\nu \sigma}-k^\nu k^\sigma \right)\epsilon ^{\mu \rho \kappa \lambda} b_\kappa k_\lambda + {\cal O}\left(\frac{m^2}{k^2}\right),
\end{eqnarray}
respectively. Then, we note that the gravitational Chern-Simons term is not generated in the case of $m\neq0$, i.e., for massive fermions. 

As we are interested in considering the finite temperature effects, we carry out the Wick rotation and split the internal momentum $p_\mu$ into its spatial and temporal components. For this, we take into account the replacements: $\eta^{\mu\nu}\to-\delta^{\mu\nu}$, i.e., $p^2\to-\delta^{\mu\nu}p^\mu p^\nu$, and so on,
\begin{equation}
\mu^{4-D}\int \frac{d^D p}{(2\pi)^D} \to \mu^{3-d}\int \frac{d^d\vec p}{(2\pi)^d} \,i\int \frac{dp_0}{2\pi},
\end{equation}
and $p^\mu\to\vec p^\mu + p_0 u^\mu$, to separate the integration variables, with $\vec p^\mu=(0,\vec p)$ and $u^\mu=(1,0,0,0)$. Besides, let us assume from now on the system to be in thermal equilibrium with a temperature $T=\beta^{-1}$, so that the antiperiodic (or periodic) boundary conditions for fermions (or bosons) lead to discrete values of $p_0=(2n+1)\frac{\pi}{\beta}$ (or $k_0=\frac{2\pi l}{\beta}$), where $n$ (or $l$) is an integer. 

Thus, by doing these considerations, after the calculation of the trace in (\ref{inta}), we arrive at:
\begin{eqnarray}
\Pi_{a}^{\mu\nu\rho\sigma} &=& \frac{\mu^{3-d}}{4\beta}\sum_{n}\int\frac{\mathrm{d}^d\vec{p}}{(2\pi)^d}\frac{\eta^{\nu\sigma}}{(\vec{p}^{\,\, 2}+p_0^2+m^2)^{2}}\nonumber\\
&\times& \left[(\vec{p}\,^2+p_0^2+m^2)\epsilon^{\mu\rho\kappa\lambda} b_\kappa k_\lambda+2\left(\vec{p}^{\,\mu}+p_0 u^\mu\right) \epsilon^{\rho\kappa\lambda\gamma} b_\kappa k_\lambda \left(\vec{p}_{\,\gamma}+p_0u_{\gamma}\right) \right.\nonumber\\
&&-2\left(\vec{p}^{\,\rho}+p_0u^\rho\right) \epsilon^{\mu\kappa\lambda\gamma} b_\kappa k_\lambda \left(\vec{p}_{\,\gamma}+p_0 u_{\gamma}\right)-2(\vec{p}\cdot \vec{k}+p_0k_0)\epsilon^{\mu\rho\kappa\lambda} b_\kappa \left(\vec{p}_\lambda+p_0 u_\lambda)\right].
\end{eqnarray}
Now, it is  convenient to perform the replacement 
\begin{equation}\label{rep}
\vec{p}^{\mu} \vec{p}^{\nu} \rightarrow \frac{\vec{p}^{\, 2}}{d}(\eta^{\mu\nu}-u^{\mu}u^{\nu}), 
\end{equation}
because of the symmetry of our integrals. By the same reasons, we can discard the odd power contributions of $\vec p_\mu$ and $p_0$, so that we can write
\begin{eqnarray}\label{intaT}
\Pi_{a}^{\mu\nu\rho\sigma} &=& \frac{\mu^{3-d}}{4d\beta}\sum_{n}\int\frac{\mathrm{d}^d\vec{p}}{(2\pi)^d}\frac{\eta^{\nu\sigma}}{(\vec{p}^{\,\, 2}+p_0^2+m^2)^{2}} \left\{[(d-6)\vec{p}\,^2+d(m^2+p_0^2)]\epsilon^{\mu\rho\kappa\lambda} k_\lambda b_{\kappa}\right.\nonumber\\
&& \left.+2(\vec{p}\,^{2}-d\, p_0^2) b_i \epsilon^{\mu \rho i \lambda} k_\lambda\right\},
\end{eqnarray}
where we have used the identity
\begin{eqnarray}\label{ident}
\eta^{\mu  \rho } \epsilon ^{\nu \sigma \alpha \lambda}-\eta^{\nu  \rho } \epsilon ^{\mu  \sigma  \alpha  \lambda}-\eta^{\lambda  \rho } \epsilon ^{\mu  \sigma \nu  \alpha}+\eta^{\rho  \sigma } \epsilon ^{\mu  \nu \alpha  \lambda }-\eta^{\alpha  \rho } \epsilon ^{\mu  \nu  \sigma \lambda} = 0,
\end{eqnarray}
to consider 
\begin{equation}
u^\mu \epsilon^{\rho\kappa\lambda\gamma} b_\kappa k_\lambda u_{\gamma}-u^\rho \epsilon^{\mu\kappa\lambda\gamma} b_\kappa k_\lambda u_{\gamma}-k_0 \epsilon^{\mu\rho\kappa\gamma} b_\kappa u_{\gamma}=-b_i \epsilon^{\mu\rho i \lambda} k_\lambda.
\end{equation}
Then, after the integration, the tensor $\Pi_{a}^{\mu\nu\rho\sigma}(k)$ takes the form
\begin{eqnarray}
\Pi_{a}^{\mu\nu\rho\sigma}(k) = A(k,m) \eta^{\nu\sigma}\epsilon^{\mu\rho\kappa\lambda} k_\lambda b_{\kappa}+B(k,m) \eta^{\nu\sigma} b_i \epsilon^{\mu \rho i \lambda} k_\lambda,
\end{eqnarray}
%\begin{eqnarray}\label{}
%\Pi_{a}^{\mu\nu\rho\sigma}(k) &=& -\frac{\Gamma\left(1-\frac{d}{2}\right)}{4\beta(4\pi)^{d/2}}\mu^{3-d}g^{\nu\sigma}\sum_{p_0}(p_0^2+m^2)^{\frac{d-4}{2}} \left\{2(p_0^2+m^2)\epsilon^{\mu\rho\kappa\lambda} b_{\kappa} k_\lambda\right.\nonumber\\
%&& \left.-[(d-1)\, p_0^2+m^2] b_i \epsilon^{\mu \rho i \lambda} k_\lambda\right\},
%\end{eqnarray}
where the coefficients $A(k,m)$ and $B(k,m)$, carrying out the dependence over the external momentum $k_\mu$, the mass $m$, and  the temperature $\beta$, in an arbitrary dimension $d$, are given by
\begin{subequations}\label{Coef1}
\begin{eqnarray}
%\label{}
A(k,m) &=& -\frac{\Gamma\left(1-\frac{d}{2}\right)}{2\beta(4\pi)^{d/2}}\mu^{3-d} \sum_{p_0}(p_0^2+m^2)^{\frac{d-2}{2}},\\
%\label{}
B(k,m) &=& \frac{\Gamma\left(1-\frac{d}{2}\right)}{4\beta(4\pi)^{d/2}}\mu^{3-d} \sum_{p_0}(p_0^2+m^2)^{\frac{d-4}{2}}[(d-1)\, p_0^2+m^2].
\end{eqnarray}
\end{subequations}

In order to do the above summations, we use the expression \cite{Ford:1979ds}
\begin{equation}\label{sum}
\sum_n\bigl[(n+b)^2+a^2\bigl]^{-\lambda} = \frac{\sqrt{\pi}\Gamma(\lambda-1/2)}{\Gamma(\lambda)(a^2)^{\lambda-1/2}}+4\sin(\pi\lambda)f_\lambda(a,b)
\end{equation}
with
\begin{equation}\label{f}
f_\lambda(a,b) = \int^{\infty}_{|a|}\frac{dz}{(z^2-a^2)^{\lambda}}Re\Biggl(\frac{1}{e^{2\pi(z+ib)}-1}\Biggl).
\end{equation}
This solution is valid only for $\lambda<1$, aside from the poles at $\lambda=1/2,-1/2,-3/2,\cdots$. However, this restriction can be circumvented if we use the recurrence relation
\begin{eqnarray}\label{rc}
f_{\lambda}(a,b) &=& -\frac1{2a^2}\frac{2\lambda-3}{\lambda-1}f_{\lambda-1}(a,b) 
- \frac1{4a^2}\frac1{(\lambda-2)(\lambda-1)}\frac{\partial^2}{\partial b^2}f_{\lambda-2}(a,b) \nonumber
\end{eqnarray}
once, twice, and so on, until $\lambda$ is placed in the range of validity. 

Then, by using the expression (\ref{sum}), the coefficients $A(k,m)$ and $B(k,m)$ can be rewritten as 
\begin{subequations}
\begin{eqnarray}
\label{A}
A(k,m) \!&=& -\frac{\sin(\frac{\pi d}{2})\mu^{3-d}}{2^2\pi^{2-d/2}\beta^{d-1}}\Gamma\left(\frac{2-d}{2}\right)\int_{|\xi|}^{\infty}d\zeta[\tanh(\pi \zeta)-1](\zeta^2-\xi^2)^{\frac{d-2}{2}}\nonumber\\
&& -\frac{(m^2)^{\frac{d-1}{2}}\mu^{3-d}}{2(4\pi)^{\frac{d+1}{2}}}\Gamma\left(\frac{1-d}{2}\right),\\
\label{B}
B(k,m) \! &=& \! \frac{\sin(\frac{\pi d}{2})\mu^{3-d}}{2^3\pi^{2-d/2}\beta^{d-1}}\Gamma\left(\frac{2-d}{2}\right)\int_{|\xi|}^{\infty}\!d\zeta\,[\tanh(\pi \zeta)-1][(d-1)\zeta^2-\xi^2](\zeta^2-\xi^2)^{\frac{d-4}{2}},\,\,\,\,\,\,\,\,\,\,\,\,
\end{eqnarray}
\end{subequations}
where $\xi=\frac{\beta m}{2\pi}$ is an adimensional parameter. Note that $A(k,m)$ has a pole in $d\rightarrow 3$, for $m \neq 0$.

Therefore, in order to cancel this singularity, as we have observed above in (\ref{PG}), and to complete the calculation,  let us focus on the tensors $\Pi_b^{\mu\nu\rho\sigma}$ and $\Pi_c^{\mu\nu\rho\sigma}$, given by Eqs.~(\ref{intb}) and (\ref{intc}), respectively. To calculate the trace over the Dirac matrices, we first use the ciclic property of the trace, to move $\gamma_5$ matrix to the end of the expression, so that finally we obtain
\begin{subequations}\label{Expartida}
\begin{eqnarray}
\Pi_{b}^{\mu\nu\rho\sigma} &=&  -\frac{\mu^{4-D}}{4}\, \int\frac{\mathrm{d}^Dp}{(2\pi)^D} \frac{(2p-k)^\nu (2p-k)^\sigma}{(p^2-m^2)^{2} (p_1^2-m^2)} \left[(p^2-m^2)\epsilon^{\mu\rho\kappa\lambda} b_\kappa (k_\lambda+p_\lambda)\right. \nonumber\\
&& \left.+2\left(p^\mu \epsilon^{\rho\kappa\lambda\tau} b_\kappa k_\lambda p_\tau-p^\rho \epsilon^{\mu\kappa\lambda\tau} b_\kappa k_\lambda p_\tau-(p\cdot k)\epsilon^{\mu\rho\kappa\lambda} b_\kappa p_\lambda\right)\right],\\
\Pi_{c}^{\mu\nu\rho\sigma} &=&  -\frac{\mu^{4-D}}{4}\, \int\frac{\mathrm{d}^Dp}{(2\pi)^D} \frac{(2p-k)^\nu (2p-k)^\sigma}{(p^2-m^2) (p_1^2-m^2)^{2}} \left[(p_1^2-m^2)\epsilon^{\mu\rho\kappa\lambda} b_\kappa (2k_\lambda-p_\lambda)\right. \nonumber\\
&& \left.+2\left(p_1^\mu \epsilon^{\rho\kappa\lambda\tau} b_\kappa k_\lambda p_\tau-p_1^\rho \epsilon^{\mu\kappa\lambda\tau} b_\kappa k_\lambda p_\tau-(p_1\cdot k)\epsilon^{\mu\rho\kappa\lambda} b_\kappa p_{1\lambda}\right)\right],
\end{eqnarray}
\end{subequations}
where we have considered $i\partial_\mu \to k_\mu$, and $p_1^\mu=p^\mu-k^\mu$. This procedure is similar to the one used in the 't Hooft-Veltman prescription \cite{tHooft:1972tcz}. By power counting, the terms $(p^2-m^2)p^{\nu} p^{\sigma}\epsilon^{\mu\rho\kappa\lambda} b_\kappa p_\lambda$ and $(p_1^2-m^2)p^{\nu} p^{\sigma}\epsilon^{\mu\rho\kappa\lambda} b_\kappa p_\lambda$ are cubically divergent. These terms are also not transversal, however, they cancel each other already in the integrand, when we consider $\Pi_b^{\mu\nu\rho\sigma}+\Pi_c^{\mu\nu\rho\sigma}$. So, the remaining divergences are at most quadratic, and thus gauge invariance is restored.

However, it is interesting to carry out the separation 
\begin{eqnarray}\label{sep}
\frac{(p^2-m^2)\epsilon^{\mu\rho\kappa\lambda} b_\kappa k_\lambda}{(p^2-m^2)^2(p_1^2-m^2)} &=& \frac{3}{2}\frac{(p^2-m^2)\epsilon^{\mu\rho\kappa\lambda} b_\kappa k_\lambda}{(p^2-m^2)^2(p_1^2-m^2)}-\frac{1}{2}\frac{(p_1^2-m^2)\epsilon^{\mu\rho\kappa\lambda} b_\kappa k_\lambda}{(p^2-m^2)(p_1^2-m^2)^2},\nonumber
\end{eqnarray}
in $\Pi_{b}^{\mu\nu\rho\sigma}$, in which we can rewrite (\ref{Expartida}), such that $\Pi_b^{\mu\nu\rho\sigma}+\Pi_c^{\mu\nu\rho\sigma}=\tilde\Pi_b^{\mu\nu\rho\sigma}+\tilde\Pi_c^{\mu\nu\rho\sigma}$, where
\begin{eqnarray}
\tilde{\Pi}_{b}^{\mu\nu\rho\sigma} &=& -\frac{\mu^{4-D}}{4}\, \int\frac{\mathrm{d}^Dp}{(2\pi)^D} \frac{(2p-k)^\nu (2p-k)^\sigma}{(p^2-m^2)^{2} (p_1^2-m^2)} \left[\frac{3}{2}(p^2-m^2)\epsilon^{\mu\rho\kappa\lambda} b_\kappa k_\lambda\right. \nonumber\\
&& \left.+2\left(p^\mu \epsilon^{\rho\kappa\lambda\tau} b_\kappa k_\lambda p_\tau-p^\rho \epsilon^{\mu\kappa\lambda\tau} b_\kappa k_\lambda p_\tau-(p\cdot k)\epsilon^{\mu\rho\kappa\lambda} b_\kappa p_\lambda\right)\right],
\end{eqnarray}
and $\tilde{\Pi}_{c}^{\mu\nu\rho\sigma}(k)=\tilde{\Pi}_{b}^{\rho\nu\mu\sigma}(-k)$. Thus, we need only to calculate $\tilde{\Pi}_{b}^{\mu\nu\rho\sigma}$, so that after using the Feynman parametrization, we have
\begin{eqnarray}\label{Pibase}
\tilde{\Pi}_{b}^{\mu\nu\rho\sigma} &=& \frac{\mu^{4-D}}{4}\, \int_{0}^{1}dx (x-1)\, \int\frac{\mathrm{d}^Dp}{(2\pi)^D} \frac{[2p^\nu+(2x-1)k^\nu] [2p^\sigma+(2x-1)k^\sigma]}{[p^2-m^2-x(x-1)k^2]^{3}}\nonumber\\
&& \times \left\{3[(p+x k)^2-m^2]\epsilon^{\mu\rho\kappa\lambda} b_\kappa k_\lambda+4\left[(p^\mu+x k^\mu) \epsilon^{\rho\kappa\lambda\tau} b_\kappa k_\lambda p_\tau \right.\right.\nonumber\\
&& \left.\left.-(p^\rho+x k^\rho) \epsilon^{\mu\kappa\lambda\tau} b_\kappa k_\lambda p_\tau-(p\cdot k+x k^2)\epsilon^{\mu\rho\kappa\lambda} b_\kappa (p_\lambda+x k_\lambda)\right]\right\}.
\end{eqnarray}
However, as we focus on finite temperature effects, let us perform the Wick rotation and split the momentum in the above expression. The result is
\begin{eqnarray}\label{Top}
\tilde{\Pi}_{b}^{\mu\nu\rho\sigma} &=& \frac{i\mu^{3-d}}{4}\, \int_{0}^{1}dx(x-1)\, \frac{1}{\beta}\sum_{n}\int\frac{\mathrm{d}^d\vec{p}}{(2\pi)^d} \left\{3[(\vec{p}+x \vec{k})^2+p_0^2+m^2]\epsilon^{\mu\rho\kappa\lambda} b_\kappa k_\lambda\right.\nonumber\\
&& \left.+4\left[\vec{p}^{\,\mu}+(p_0-xk_0)u^\mu+x k^\mu\right] \epsilon^{\rho\kappa\lambda\gamma} b_\kappa k_\lambda \left[\vec{p}_{\,\gamma}+(p_0-xk_0)u_{\gamma}\right] \right.\nonumber\\
&& \left.-4\left[\vec{p}^{\,\rho}+(p_0-xk_0)u^\rho+x k^\rho\right] \epsilon^{\mu\kappa\lambda\gamma} b_\kappa k_\lambda \left[\vec{p}_{\,\gamma}+(p_0-xk_0)u_{\gamma}\right] \right.\nonumber\\
&& \left.-4[\vec{p}\cdot \vec{k}+(p_0-xk_0)k_0+x k^2]\epsilon^{\mu\rho\kappa\lambda} b_\kappa \left[\vec{p}_\lambda+(p_0-xk_0)u_\lambda+x k_\lambda)\right]\right\}\nonumber\\
&& \times \frac{[2\vec{p}^{\, \nu}+(2x-1)k^\nu+2(p_0-xk_0)u^\nu] [2\vec{p}^{\,\sigma}+(2x-1)k^\sigma+2(p_0-xk_0) u^\sigma]}{[\vec{p}^{\,\, 2}+(p_0-x k_0)^2+m^2+x(1-x)k^2]^{3}}.
\end{eqnarray}

Now, in order to select the tensorial structures of the above result (\ref{Top}), let us consider the expressions (\ref{rep}) and 
\begin{eqnarray}
\vec{p}^{\, \mu} \vec{p}^{\, \nu} \vec{p}^{\, \alpha} \vec{p}^{\, \beta} &\rightarrow& \frac{\vec{p}^{\, 4}}{d(d+2)}\left[(\eta^{\mu\nu}-u^{\mu}u^{\nu})(\eta^{\alpha\beta}-u^{\alpha}u^{\beta})+(\eta^{\mu\alpha}-u^{\mu}u^{\alpha})(\eta^{\nu\beta}-u^{\nu}u^{\beta})\right.\nonumber\\
&& \left.+(\eta^{\mu\beta}-u^{\mu}u^{\beta})(\eta^{\alpha\nu}-u^{\alpha}u^{\nu})\right],
\end{eqnarray}
as well as the identity (\ref{ident}), so that we obtain
\begin{eqnarray}\label{Pib}
\tilde{\Pi}_{b}^{\mu\nu\rho\sigma} &=& C(k,m) \eta^{\nu\sigma} \epsilon^{\mu\rho \kappa \lambda} b_\kappa k_\lambda -D(k,m) \eta^{\nu\sigma}b_i\epsilon^{\mu\rho i \lambda} k_\lambda+ E(k,m) k^{\nu} k^{\sigma} \epsilon^{\mu\rho \kappa \lambda} b_\kappa k_\lambda\nonumber\\
&& -F(k,m) k^{\nu}k^{\sigma}b_i\epsilon^{\mu\rho i \lambda} k_\lambda+ \left[G(k,m)-2 L(k,m)\right] (k^{\nu} u^{\sigma}+k^{\sigma}u^{\nu}) \epsilon^{\mu\rho \kappa \lambda} b_\kappa k_\lambda\nonumber\\
&& +\left[H(k,m)-4D(k,m)\right] u^{\nu}u^{\sigma}\epsilon^{\mu\rho \kappa \lambda} b_\kappa k_\lambda-I(k,m) (k^{\nu}u^{\sigma}+u^{\nu}k^{\sigma})b_i\epsilon^{\mu\rho i \lambda} k_\lambda\nonumber\\
&&-J(k,m) u^{\nu} u^{\sigma} b_i \epsilon^{\mu\rho i \lambda} k_\lambda + K(k,m) (b^{\nu} \epsilon ^{\mu \rho \sigma \lambda} k_\lambda+b^{\sigma} \epsilon ^{\mu \rho \nu \lambda} k_\lambda ) \nonumber\\
&& + D(k,m)[-(u^\nu b^{\sigma}+u^\sigma b^{\nu}) \epsilon ^{\mu \rho \lambda \tau} k_\lambda u_\tau+b_{0}( u^{\nu}  \epsilon ^{\mu \rho \sigma \lambda} k_\lambda+ u^{\sigma} \epsilon ^{\mu \rho \nu \lambda} k_\lambda)]\nonumber\\
&& +L(k,m) [-(k^\nu b^{\sigma}+k^\sigma b^{\nu}) \epsilon ^{\mu \rho \lambda \tau} k_\lambda u_\tau+b_{0}( k^{\nu}  \epsilon ^{\mu \rho \sigma \lambda} k_\lambda+ k^{\sigma} \epsilon ^{\mu \rho \nu \lambda} k_\lambda) ].
\end{eqnarray}
with
\begin{subequations}\label{Coef2}
\begin{eqnarray}\label{C}
C(k,m) &=& \frac{\Gamma\left(1-\frac{d}{2}\right)\mu^{3-d}}{4\beta(4\pi)^{d/2}} \sum_{p_0}\int_0^1 dx\,(1-x)\left[(p_0-x k_0)^2+M^2\right]^{\frac{d-4}{2}} \\
&& \times \left\{4\left[(p_0-x k_0)^2+M^2\right]+(2-d) x k^2\right\},\nonumber
\end{eqnarray}
\begin{eqnarray}
D(k,m) &=& \frac{\Gamma\left(1-\frac{d}{2}\right)\mu^{3-d}}{2\beta(4\pi)^{d/2}} \sum_{p_0}\int_0^1 dx\,(1-x) \left[(p_0-x k_0)^2+M^2\right]^{\frac{d-4}{2}} \nonumber\\
&&\times \left[(d-1) (p_0-x k_0)^2+M^2\right],
\end{eqnarray}
\begin{eqnarray}
E(k,m) &=& \frac{\Gamma \left(3-\frac{d}{2}\right)\mu^{3-d}}{4\beta(4\pi)^{d/2}} \sum_{p_0}\int_0^1 dx\, (1-2 x)^2 M^2 \left[(p_0-x k_0)^2+M^2\right]^{\frac{d-6}{2}},
 \end{eqnarray}
\begin{eqnarray}
F(k,m) &=& \frac{\Gamma \left(2-\frac{d}{2}\right)\mu^{3-d}}{4\beta(4\pi)^{d/2}} \sum_{p_0}\int_0^1 dx\, (1-x)(1-2 x)^2 \left[(p_0-x k_0)^2+M^2\right]^{\frac{d-6}{2}} \nonumber\\
&&\times \left[(d-3) (p_0-x k_0)^2+M^2\right],
\end{eqnarray}
\begin{eqnarray}
G(k,m) &=& \frac{\Gamma \left(3-\frac{d}{2}\right)\mu^{3-d}}{2\beta(4\pi)^{d/2}}k^2 \sum_{p_0}\int_0^1 dx\,(1-x) x (2x-1)(p_0-x k_0) \left[(p_0-x k_0)^2+M^2\right]^{\frac{d-6}{2}},\nonumber
\end{eqnarray}
\begin{eqnarray}
H(k,m) &=& \frac{-\Gamma \left(2-\frac{d}{2}\right)\mu^{3-d}}{2\beta(4\pi)^{d/2}}k^2 \sum_{p_0}\int_0^1dx\,(1-x)x\left[(p_0-x k_0)^2+M^2\right]^{\frac{d-6}{2}} \nonumber\\
&&\times \left[(d-3)(p_0-x k_0)^2+M^2\right],
\end{eqnarray}
\begin{eqnarray}
I(k,m) &=& \frac{\Gamma \left(2-\frac{d}{2}\right)\mu^{3-d}}{2\beta(4\pi)^{d/2}} \sum_{p_0}\int_0^1 dx\,(1-x) (1-2x)(p_0-x k_0) \left[(p_0-x k_0)^2+M^2\right]^{\frac{d-6}{2}} \nonumber\\
&&\times \left[(d-1)(p_0-x k_0)^2+3M^2\right],
\end{eqnarray}
\begin{eqnarray}
J(k,m) &=& \frac{\mu^{3-d}}{2\beta(4\pi)^{d/2}}\sum_{p_0}\int_0^1dx\,(x-1) \left\{3 \Gamma \left(1-\frac{d}{2}\right)\left[(p_0-x k_0)^2+M^2\right]^{\frac{d}{2}-1}\right.\nonumber\\
&& \left.-12 \Gamma \left(2-\frac{d}{2}\right) (p_0-x k_0)^2 \left[(p_0-x k_0)^2+M^2\right]^{\frac{d}{2}-2}\right.\nonumber\\
&&\left.+4 \Gamma \left(3-\frac{d}{2}\right) (p_0-x k_0)^4 \left[(p_0-x k_0)^2+M^2\right]^{\frac{d}{2}-3}\right\},
\end{eqnarray}
\begin{eqnarray}
K(k,m) &=& \frac{\Gamma \left(1-\frac{d}{2}\right)\mu^{3-d}}{2\beta(4\pi)^{d/2}} \sum_{p_0}\int_0^1 dx\,(1-x) \left[(p_0-x k_0)^2+M^2\right]^{\frac{d-2}{2}},
\end{eqnarray}
\begin{eqnarray}
L(k,m) &=& \frac{\Gamma \left(2-\frac{d}{2}\right)\mu^{3-d}}{2\beta(4\pi)^{d/2}} \sum_{p_0}\int_0^1 dx\,(1-x) (1-2x)(p_0-x k_0) \left[(p_0-x k_0)^2+M^2\right]^{\frac{d-4}{2}},
\end{eqnarray}
\end{subequations}
where $M^2=m^2-x(x-1)k^2$. Note that, in the zero temperature and mass limit, i.e., if the summation over integer $n$ is replaced by the integral $\frac{1}{2\pi}\int dp_0$ and $m\to0$, we obtain $C=-\frac{k^2}{192 \pi ^2}=-Ek^2$ and $D=0=F=G=H$, which matches the known zero-temperature result (\ref{Pim0}) \cite{Mariz:2004cv}. In order to further simplify the above equations, we can make the change of variables $p_0\to p_0+xk_0$ (actually, one should, first, do this change of variables, and then, introduce the discrete $p_0$), which allows to rule out the dependence on $k_0$, except in $M$.

Now, by manipulating these expressions (\ref{Coef2}), we found the relations
\begin{subequations}
\begin{eqnarray}
k^2 G(k,m)+k_0 H(k,m)&=&2k_0 D(k,m),\\
k^2 I(k,m)+k_0 J(k,m)&=&0,\\
k^2 L(k,m)+k_0 D(k,m)&=&0,\\
k^2 F(k,m)+k_0 I(k,m)&=&-D(k,m)+B(k,m),\\
C(k,m)+k^2 E(k,m)+k_0 G(k,m)&=& 2k_0 L(k,m)-A(k,m),
\end{eqnarray}
\end{subequations}
where the functions $A(k,m)$ and $B(k,m)$ were already defined in (\ref{Coef1}). From these relations, the tensor~(\ref{Pib}) can be rewritten as follows:
\begin{eqnarray}\label{Pib2}
\tilde{\Pi}_{b}^{\mu\nu\rho\sigma} &=& C(k,m) \left(\eta^{\nu\sigma}-\frac{k^{\nu} k^{\sigma}}{k^2}\right) \epsilon^{\mu\rho \kappa \lambda} b_\kappa k_\lambda -D(k,m) \left(\eta^{\nu\sigma}-\frac{k^{\nu} k^{\sigma}}{k^2}\right)b_i\epsilon^{\mu\rho i \lambda} k_\lambda\nonumber\\ 
&&-A(k,m)\frac{k^{\nu}k^{\sigma}}{k^2}\epsilon^{\mu \rho \kappa \lambda} b_\kappa k_\lambda - B(k,m)\frac{k^{\nu}k^{\sigma}}{k^2}b_i \epsilon^{\mu \rho i \lambda} k_\lambda\nonumber\\
&&+ \left[H(k,m)-4D(k,m)\right]\left(\frac{k_0}{k^2}k^{\nu}-u^{\nu}\right)\left(\frac{k_0}{k^2}k^{\sigma}-u^{\sigma}\right) \epsilon^{\mu\rho \kappa \lambda} b_\kappa k_\lambda \nonumber\\
&&- J(k,m)\left(\frac{k_0}{k^2}k^{\nu}-u^{\nu}\right)\left(\frac{k_0}{k^2}k^{\sigma}-u^{\sigma}\right) b_i\epsilon^{\mu \rho i \lambda} k_\lambda+ K(k,m) (b^{\nu} \epsilon ^{\mu \rho \sigma \lambda} k_\lambda+b^{\sigma} \epsilon ^{\mu \rho \nu \lambda} k_\lambda )\nonumber\\
&&+ D(k,m)[-(u^\nu b^{\sigma}+u^\sigma b^{\nu}) \epsilon ^{\mu \rho \lambda \tau} k_\lambda u_\tau+b_{0}( u^{\nu}  \epsilon ^{\mu \rho \sigma \lambda} k_\lambda+ u^{\sigma} \epsilon ^{\mu \rho \nu \lambda} k_\lambda) ] \nonumber\\
&&-D(k,m)\frac{k_0}{k^2} [-(k^\nu b^{\sigma}+k^\sigma b^{\nu}) \epsilon ^{\mu \rho \lambda \tau} k_\lambda u_\tau+b_{0}( k^{\nu} \epsilon^{\mu \rho \sigma \lambda} k_\lambda+ k^{\sigma} \epsilon ^{\mu \rho \nu \lambda} k_\lambda) ],
\end{eqnarray}
where the structures $ \left(g^{\nu\sigma}-\frac{k^{\nu} k^{\sigma}}{k^2}\right)$ and $\left(\frac{k_0}{k^2}k^{\nu}-u^{\nu}\right)$ are gauge invariant.  In fact, only the mass dependent term, $\frac{k^{\nu}k^{\sigma}}{k^2}$, is not gauge invariant, however, when we collect all contributions to the vacuum polarization tensor, given by $\Pi^{\mu\nu\rho\sigma}=\Pi_a^{\mu\nu\rho\sigma}+\frac12\Pi_b^{\mu\nu\rho\sigma}+\frac12\Pi_c^{\mu\nu\rho\sigma} = \Pi_a^{\mu\nu\rho\sigma}+\tilde{\Pi}_b^{\mu\nu\rho\sigma}$, we can easily observed the gauge invariance in $\Pi^{\mu\nu\rho\sigma}$. Thus, by dropping the last three terms, which evidently do not contribute to the gravitational Chern-Simons term, we have 
\begin{eqnarray}
\label{Pib3}
\Pi^{\mu\nu\rho\sigma} &=& \left[A(k,m)+C(k,m)\right] \left(\eta^{\nu\sigma}-\frac{k^{\nu} k^{\sigma}}{k^2}\right) \epsilon^{\mu\rho \kappa \lambda} b_\kappa k_\lambda \nonumber\\ 
&&+ \left[B(k,m)-D(k,m)\right] \left(\eta^{\nu\sigma}-\frac{k^{\nu} k^{\sigma}}{k^2}\right)b_i\epsilon^{\mu\rho i \lambda} k_\lambda\nonumber\\
&& -\left[4D(k,m)-H(k,m)\right]\left(\frac{k_0}{k^2}k^{\nu}-u^{\nu}\right)\left(\frac{k_0}{k^2}k^{\sigma}-u^{\sigma}\right) \epsilon^{\mu\rho \kappa \lambda} b_\kappa k_\lambda \nonumber\\
&& -J(k,m)\left(\frac{k_0}{k^2}k^{\nu}-u^{\nu}\right)\left(\frac{k_0}{k^2}k^{\sigma}-u^{\sigma}\right) b_i\epsilon^{\mu \rho i \lambda} k_\lambda.
\end{eqnarray}
In the above coefficients, if we return to the $p_0$ integral, i.e., $\frac{1}{\beta}\sum_{p_0} \to \frac{1}{2\pi}\int dp_0$, after the calculation, we obtain exactly the result (\ref{PG}), as expected. We note that, unlike the previous papers \cite{Landsteiner:2012kd,Chernodub:2013kya,Yee:2014dxa,Chowdhury:2015pba}, where massless chiral fermions were considered, we carry out our calculations for a completely arbitrary spinor.

Now, let us show explicitly the cancellation of the singularity between the coefficients $A(k,m)$ and $C(k,m)$, as implicitly observed in (\ref{PG}). As the divergent term is only mass-dependent (see second term of~(\ref{A})), we first take into account $k_0=0$ in Eq.~(\ref{C}), and then, by evaluating the summation over $p_0$, we obtain
%\textbf{\textit{As was already seen, the singularity rest on the mass dependence and, for this, it suffices to single out this dependence doing $k_\mu \to 0$ throught (by means of) the static limit $(k_0= 0, \vec{k}\rightarrow 0)$}} \cite{Assuncao:2016fko}. Then, by first evaluating the summation of Eq.~(\ref{C}), we obtain,
\begin{eqnarray}
\!\!\!\! C(k,m)|_{k_0=0} &=& \frac{\Gamma \left(\frac{1-d}{2}\right)\mu^{3-d}}{4(4\pi) ^{\frac{d+1}{2}}} \int_0^{1}dx \left(M^2\right)^{\frac{d-3}{2}} \left[(3-d) M^2+(d-1)m^2\right] + \frac{\sin \left(\frac{\pi  d}{2}\right)\Gamma \left(\frac{2-d}{2}\right)\mu^{3-d}}{8\pi ^{2-\frac{d}{2}}\beta^{d-1}} \nonumber\\
&&\times\int_{|\xi'|}^{\infty}d\zeta\, [\tanh (\pi \zeta)-1] (\zeta^2-\xi'^2)^{\frac{d}{2}-2} [2(\zeta^2-\xi'^2)+(d-2)(\xi'^2-\xi^2)],
\end{eqnarray}
where $\xi'=\frac{\beta M}{2\pi}$. Now, in order to single out the  pole part ($PP$) of the above expression, let us finally consider $\vec{k} \to 0$, so that we get
\begin{eqnarray}
PP[C(0,m)] &=& \frac{\left(m^2\right)^{\frac{d-1}{2}}\mu^{3-d}}{2(4\pi) ^{\frac{d+1}{2}}} \Gamma \left(\frac{1-d}{2}\right),
\end{eqnarray}
which precisely cancels the pole part of the coefficient $A(0,m)$. Therefore, we are seeing that $\Pi^{\mu\nu\rho\sigma}$ (\ref{Pib3}) is finite even when the mass is taken into account.

Our next step is to calculate the coefficients (\ref{Coef2}), by considering, in the coefficients accompanying $k^2$ and $T^2$, the static limit $(k_0= 0, \vec{k}\rightarrow 0)$, which is the one used to obtain the anomalous conductivities (i.e., the chiral vortical conductivity) in Weyl semimetals  ($m=0$) \cite{Landsteiner:2013sja}. Furthermore, if we do not take into account these conditions, additional term with five and more derivatives will arise. Then, by using these considerations, we can evaluate the coefficients (\ref{Coef2}), as follows:
\begin{subequations}\label{Coef2T}
\begin{eqnarray}
A \!&=& -\frac{\sin(\frac{\pi d}{2})\mu^{3-d}}{2^2\pi^{2-d/2}\beta^{d-1}}\Gamma\left(\frac{2-d}{2}\right)\int_{0}^{\infty}d\zeta[\tanh(\pi \zeta)-1]\zeta^{d-2}\nonumber\\
&=& \frac{T^2}{48},
\end{eqnarray}
\begin{eqnarray}
B \! &=& \! (d-1)\frac{\sin(\frac{\pi d}{2})\mu^{3-d}}{2^3\pi^{2-d/2}\beta^{d-1}}\Gamma\left(\frac{2-d}{2}\right)\int_{0}^{\infty}\!d\zeta\,[\tanh(\pi \zeta)-1]	\zeta^{d-2}.\nonumber\\
&=& -\frac{T^2}{48},
\end{eqnarray}
\begin{eqnarray}
C &=& \frac{\Gamma \left(\frac{1}{2}-\frac{d}{2}\right)\mu^{3-d}}{4(4\pi)^{(d+1)/2}} \int_0^1 dx\, (5-4 x-d)  \left(M^2\right)^{\frac{d-1}{2}}+\frac{\beta^{1-d}\Gamma \left(1-\frac{d}{2}\right)\mu^{3-d}}{2\pi ^{(4-d)/2}} \sin \left(\frac{\pi  d}{2}\right) \nonumber\\
&&\times \int_0^1 dx\,(1-x) \int_{0}^{\infty}d\zeta\, \left[\tanh (\pi  \zeta)-1\right] \left(\zeta^2\right)^{\frac{d}{2}-1} \nonumber\\
&=& -\frac{k^2}{192 \pi ^2}-\frac{T^2}{48},
\end{eqnarray}
\begin{eqnarray}
D &=& \frac{\beta^{1-d}\Gamma \left(1-\frac{d}{2}\right)\mu^{3-d}}{4\pi ^{2-\frac{d}{2}}}(d-1) \sin \left(\frac{\pi  d}{2}\right) \int_0^1 dx\,(1-x)\, \int_{0}^{\infty}d\zeta\, (\tanh (\pi  \zeta)-1) \left(\zeta^2\right)^{\frac{d}{2}-1} \nonumber\\
&=& -\frac{T^2}{48},
\end{eqnarray}
\begin{eqnarray}
H &=& \frac{\Gamma \left(1-\frac{d}{2}\right)\mu^{3-d}}{8(4\pi)^{\frac{d}{2}-1}} \sin \left(\frac{\pi  d}{2}\right) \left(\frac{\beta}{2\pi}\right)^{3-d}k^2\int_0^1 dx\,x(1-x)\int_{0}^\infty\, d\zeta\, \zeta^{d-2}\tanh (\pi  \zeta) \text{sech}^2(\pi  \zeta) \nonumber\\
&=& 0,
\end{eqnarray}
\begin{eqnarray}
J &=& \frac{\beta^{3-d}\Gamma \left(1-\frac{d}{2}\right)\mu^{3-d}}{8\pi ^{2-\frac{d}{2}}} \sin \left(\frac{\pi  d}{2}\right)k^2\int_0^1dx\,(1-x)^2\,x\,\int_{0}^{\infty}d\zeta\, \tanh (\pi  \zeta) \text{sech}^2(\pi  \zeta) \zeta^{d-2} \nonumber\\
&& +\frac{(1-d^2)\beta^{1-d}\Gamma \left(1-\frac{d}{2}\right)\mu^{3-d}}{4\pi ^{2-\frac{d}{2}}}\sin \left(\frac{\pi  d}{2}\right) \int_0^1 dx\,(1-x)\int_{0}^{\infty}d\zeta\, (\tanh (\pi  \zeta)-1) \zeta^{d-2} \nonumber\\
&=& \frac{T^2}{12}.
\end{eqnarray}
\end{subequations}
Thus, for Eq.~(\ref{Pib3}), we obtain
\begin{eqnarray}\label{Pilast}
\Pi^{\mu\nu\rho\sigma} &=& -\frac{k^2}{192 \pi ^2} \left(\eta^{\nu\sigma}-\frac{k^{\nu} k^{\sigma}}{k^2}\right) \epsilon^{\mu\rho \kappa \lambda} b_\kappa k_\lambda +\frac{T^2}{12}\left(\frac{k_0}{k^2}k^{\nu}-u^{\nu}\right)\left(\frac{k_0}{k^2}k^{\sigma}-u^{\sigma}\right)b_0 \epsilon^{\mu\rho \kappa \lambda} u_\kappa k_\lambda.\,\,\,\,\,
\end{eqnarray}
We note that the above term $u^{\nu}u^{\sigma}b_0 \epsilon^{\mu\rho \kappa \lambda} u_\kappa k_\lambda$, with the coefficient $\frac{T^2}{12}$, is the one that contributes to the chiral vortical conductivity, which was obtained in Refs.~\cite{Yee:2014dxa,Chowdhury:2015pba}, in the context of Kubo formulation and derivative expansion. However, as we can see, our result (\ref{Pilast}) is a general one, gauge invariant, and finite.  

By returning to the coordinate space, through the replacement $k^\mu \rightarrow i \partial^\mu$, we have
\begin{eqnarray}
\label{result}
S_\RM{CS}[h] &=&  \int d^4x\, h_{\mu\nu} \left[-\frac{1}{192 \pi ^2}\epsilon^{\mu \rho \kappa \lambda} b_\kappa \partial_\lambda \left(\Box h_{\rho}\,\!^{\nu}-\partial^{\nu} \partial^{\sigma} h_{\rho\sigma}\right)\right. \nonumber\\
&& \left. -\frac{T^2}{12} b_0 \epsilon^{\mu\rho \kappa \lambda} u_\kappa \partial_\lambda \left(\frac{\partial_0 \partial^{\nu}}{\Box} -u^{\nu}\right)\left(\frac{\partial_0 \partial^{\sigma}}{\Box} -u^{\sigma} \right)h_{\rho\sigma}\right].
\end{eqnarray}
We note that the nonlocality of our result is a price we pay for the gauge invariance. The similar situation was found to occur for the one-derivative term in \cite{Ferrari:2006gs}. 

It is instructive to compare the four-dimensional gravitational Chern-Simons term with the two-dimensional one. As it is well known, see f.e. \cite{Gur},  the two-dimensional gravitational Chern-Simons term is known to have the form
\begin{eqnarray}
S_{2D}=-\frac{1}{8\pi^2}\int d^2x \sqrt{|g|}(fr+f^3)=\frac{1}{4\pi}\int d^2x \Theta f,
\end{eqnarray}
where $f=\epsilon^{\mu\nu}f_{\mu\nu}$, $\epsilon^{01}=1$, and $f_{\mu\nu}=\partial_{\mu} a_{\nu}-\partial_{\nu} a_{\mu}=\sqrt{|g|}\epsilon_{\mu\nu}f$.
The $\Theta$ is the Chern-Simons coefficient.

Let us proceed along the lines developed in \cite{Jackproj}. First, we choose $\Theta=x^{\alpha}b_{\alpha}$, where $b_{\alpha}$ is a constant vector implementing the Lorentz symmetry breaking. Then, we have
\begin{eqnarray}
S_{2D}=\frac{1}{2\pi}\int d^2x x^{\alpha}b_{\alpha} \epsilon^{\mu\nu}\partial_{\mu} a_{\nu}.
\end{eqnarray}
 Now, we integrate by parts and disregard the superficial term, so that we arrive at
\begin{eqnarray}
S_{2D}=-\frac{1}{2\pi}b_{\mu}\epsilon^{\mu\nu}A_{\nu}.
\end{eqnarray}
Here, we have the new vector $A_{\nu}=\int d^2x a_{\nu}$. It is constant since the coordinate dependence is integrated out. However, in our theory there is only one privileged direction, that is, the $b_{\nu}$ vector, which requires $A_{\nu}=\lambda b_{\nu}$. But in this case our Chern-Simons term identically disappears, so, for the Lorentz symmetry breaking of this form, the two-dimensional gravitational Chern-Simons term is equal to zero. This result can be confirmed by direct calculations of Eq.~(\ref{CS2}), in two dimensions, as well. The Eq.~(\ref{CS1}) is not considered, because in two dimensions the spin connection $\omega_\mu$ vanishes.

Let us discuss our results. We proved that the four-dimensional gravitational Chern-Simons term turns out to be finite (\ref{PG}), despite the initial expression for it strongly (cubically) diverges. We also found that differently from the previous papers studying gravitational anomalies at finite temperature \cite{Landsteiner:2012kd,Chernodub:2013kya,Yee:2014dxa,Chowdhury:2015pba}, our result (\ref{Pib3}) is obtained for the presence of one fermion only, and it is generic, without any restrictions for the spinor fields, while in these references the spinors are suggested to be massless and chiral. Differently from these papers, we started from the action with the explicit Lorentz symmetry breaking, considered the coupling of fermion to the gravity only, and carried all calculations explicitly with use of the Matsubara frequencies methodology, without use the derivative expansion. Also, we noted  that our result (\ref{Pilast}) for the gravitational Chern-Simons term was obtained in the zero mass limit, however, for the non-zero mass case the gauge symmetry is observed as well (see Eq.~(\ref{Pib3})). Then, we can conclude that our result is consistent with the previous studies. It is necessary to note that, unlike many other finite temperature studies, our result displays very simple temperature dependence, which moreover monotonously grows with the temperature. This can indicate that our theory is probably an effective one for the low-temperature regime, while the whole temperature range should be described by a more involved theory.

\textbf{Acknowledgments.}  This work by A. Yu. P. was partially supported by Conselho Nacional
de Desenvolvimento Científico e Tecnológico (CNPq), project No. 303783/2015-0.

\end{document}